\documentclass[aps,amsfonts,floatfix,reprint,tightenlines,amssymb,superscriptaddress,twocolumn,longbibliography]{revtex4-2}  

\usepackage{amsmath}
\usepackage{subcaption}
\usepackage{enumerate}

\usepackage[linktocpage=true,
  colorlinks=true, 
  pdfborder={0 0 0},
  linkcolor=blue,
  citecolor=red,
  filecolor=yellow,
  urlcolor=blue,
  bookmarks,
]{hyperref}
\usepackage[capitalize]{cleveref}

\usepackage{orcidlink}

\newcommand{\angstrom}{\textup{\AA}}
\begin{document}

\preprint{APS/123-QED}

\title{Machine learning supported annealing for prediction of grand canonical crystal
structures}

\author{Yannick Couzini\'{e}\orcidlink{0000-0002-5408-8197}}
\email{couzinie.y.aa@m.titech.ac.jp}
\affiliation{
        Department of Physics,
        The University of Tokyo
        Hongo, Bunkyo-ku,
        Tokyo, Japan.
}
\affiliation{
Quemix Inc.,
Taiyo Life Nihombashi Building,
2-11-2,
Nihombashi Chuo-ku, 
Tokyo,
Japan
}

\author{Yuya Seki\orcidlink{0000-0001-5393-2280}}
\affiliation{
Graduate School of Science and Technology,
Keio University,
3-14-1 Hiyoshi, Kohoku-ku, Yokohama,
Kanagawa,
Japan
}

\affiliation{
        Keio University Sustainable Quantum Artificial Intelligence Center (KSQAIC),
        Keio University,
        2-15-45 Mita, Minato-ku, Tokyo,
        Japan
}

\author{Yusuke Nishiya \orcidlink{0000-0001-6526-0936}}
\affiliation{
        Department of Physics,
        The University of Tokyo
        Hongo, Bunkyo-ku,
        Tokyo, Japan.
}

\affiliation{
Quemix Inc.,
Taiyo Life Nihombashi Building,
2-11-2,
Nihombashi Chuo-ku, 
Tokyo,
Japan
}

\author{Hirofumi Nishi\orcidlink{0000-0001-5155-6605}}
\affiliation{
        Department of Physics,
        The University of Tokyo
        Hongo, Bunkyo-ku,
        Tokyo, Japan.
}

\affiliation{
Quemix Inc.,
Taiyo Life Nihombashi Building,
2-11-2,
Nihombashi Chuo-ku, 
Tokyo,
Japan
}

\author{Taichi Kosugi\orcidlink{0000-0003-3379-3361}}
\affiliation{
        Department of Physics,
        The University of Tokyo
        Hongo, Bunkyo-ku,
        Tokyo, Japan.
}

\affiliation{
Quemix Inc.,
Taiyo Life Nihombashi Building,
2-11-2,
Nihombashi Chuo-ku, 
Tokyo,
Japan
}

\author{Shu Tanaka\orcidlink{0000-0002-0871-3836}}
\affiliation{
        Keio University Sustainable Quantum Artificial Intelligence Center (KSQAIC),
        Keio University,
        2-15-45 Mita, Minato-ku, Tokyo,
        Japan
}
\affiliation{
Department of Applied Physics and Physico-Informatics,
Keio University,
3-14-1 Hiyoshi, Kohoku-ku, Yokohama,
Kanagawa,
Japan
}
\affiliation{
        Human Biology-Microbiome-Quantum Research Center (WPI-Bio2Q),
        Keio University,
        2-15-45 Mita, Minato-ku, Tokyo,
}
\affiliation{
        Green Computing Systems Research Organization (GCS),
        Waseda University,
        162-0042 Wasedamachi, Shinjuku-ku, Tokyo,
        Japan
}

\author{Yu-ichiro Matsushita\orcidlink{0000-0002-9254-5918}}
\affiliation{
        Department of Physics,
        The University of Tokyo
        Hongo, Bunkyo-ku,
        Tokyo, Japan.
}
\affiliation{
Quemix Inc.,
Taiyo Life Nihombashi Building,
2-11-2,
Nihombashi Chuo-ku, 
Tokyo,
Japan
}
\affiliation{
Quantum Material and Applications Research Center,
National Institutes for Quantum Science and Technology (QST),
2-12-1, Ookayama, Meguro-ku, Tokyo, Japan
}

\date{\today}

\begin{abstract}
        This study investigates the application of Factorization Machines with
        Quantum Annealing (FMQA) to address the crystal structure problem (CSP)
        in materials science. FMQA is a black-box optimization algorithm that
        combines machine learning with annealing machines to find samples to a
        black-box function that minimize a given loss. The CSP involves
        determining the optimal arrangement of atoms in a material based on its
        chemical composition, a critical challenge in materials science. We
        explore FMQA's ability to efficiently sample optimal crystal
        configurations by setting the loss function to the energy of the
        crystal configuration as given by a predefined interatomic potential.
        Further we investigate how well the energies of the various metastable
        configurations, or local minima of the potential, are learned by the
        algorithm. Our investigation reveals FMQA's potential in quick ground
        state sampling and in recovering relational order between local minima.
\end{abstract}

\maketitle


\section{\label{sec:intro} Introduction}
Solving the crystal structure problem (CSP) from chemical composition remains an enduring challenge in
materials science, demanding innovative methodologies to overcome the
exponential growth of potential structures as the system size
increases~\cite{maddox:1988aa}. Various approaches on classical computers have
been developed~\cite{oganov:2019aa}, e.g.\ random
search~\cite{pickard:2006aa,pickard:2007aa,pickard:2011aa,needs:2016aa},
simulated annealing (SA)~\cite{wille:1987aa, wille:2000aa, yin:2022aa}, minima
hopping~\cite{goedecker:2004aa,amsler:2010aa}, evolutionary
algorithms~\cite{bush:1995aa,oganov:2006aa,oganov:2011aa,lyakhov:2013aa} and
particle swarm optimization~\cite{wang:2010aa,zhang:2017aa}. Various software
suites such as USPEX~\cite{glass:2006aa,oganov:2011ab},
CALYPSO~\cite{wang:2010aa,wang:2012aa,wang:2014aa}, and
CrySPY~\cite{yamashita:2021aa} are used in standard approaches to the CSP.
Being on classical hardware, these algorithms do not manage to compute large
scale structures. With the continually increasing computational capacity of
quantum computers, there is a need to develop efficient algorithms to solve the
CSP that is compatible with quantum hardware.

In recent years, the use of quantum computers has attracted a great deal of
attention as a means of searching for globally optimal
solutions~\cite{kadowaki:1998aa,durr1999quantum,tanaka:2017aa,albash:2018aa,jones:2019aa,mcardle:2019aa,kosugi:2022ab}.
Quantum computers are characterized by their ability to escape from locally
stable solutions and accelerate the search for globally optimal solutions by
utilizing the quantum tunneling effect~\cite{denchev:2016aa,albash:2018aa} and
the superposition of states to concurrently evaluate a large state
space~\cite{nielsen:2001aa}. Quantum annealing (QA)
machines~\cite{kadowaki:1998aa, hauke:2020aa, king:2022aa,
catherine:2019aa,boothby:2020aa} and gate-based quantum
computers~\cite{nielsen:2001aa} are the two
main current architectures in development. Exhaustive structure search schemes
using gate-based quantum
computers~\cite{hirai:2022aa,kosugi:2023aa,nishiya:2024aa}, quantum
annealers and the equivalent Ising
machines~\cite{gusev:2023aa,couzinie:2024aa,ichikawa:2023aa,nawa:2023aa,sampei:2023aa}
have recently been proposed.

Using the encoding proposed in~\cite{couzinie:2024aa} it is possible to encode
the local interactions of any potential with a hard cutoff into a quadratic
unconstrained binary optimization (QUBO) or higher order
unconstrained binary optimization (HUBO) problem by discretizing the simulation
cell. Optimizing the structure of ionically bonded crystals can then be
described as a QUBO problem (see~\cite{gusev:2023aa}) and that of
covalently bounded crystals can be described by a third-order HUBO problem
(see~\cite{couzinie:2024aa}). When it comes to non-crystalline phases of
covalently bounded materials or metals, more accurate potentials such as
bond-order~\cite{tersoff:1988aa,finnis:1984aa}, machine
learned~\cite{botu:2017aa} or embedded atom method (EAM) based
potentials~\cite{daw:1984aa,lee:2010aa} are needed. In these cases the order of
the HUBO is generally equal to the amount of neighbouring lattice sites inside
the cutoff radius of the respective potential.

Current Ising machines~\cite{mohseni:2022aa} such as quantum annealing
hardware~\cite{boothby:2020aa}, or classical annealers
(e.g.~\cite{aramon:2019aa}) usually focus on the solving of QUBO problems.
There are exceptions for higher order such as the QAOA
algorithm~\cite{farhi:2014aa} or simulated bifurcation
machines~\cite{kanao:2022aa}, but performance in these cases still suffer from
the quickly increasing number of non-zero interactions. It is possible to
reduce HUBO problems to QUBO problems through usual quadratization
techniques~\cite{dattani:2019aa}, but these usually add a number of auxiliary
bits in the same order of the number of removed non-zero interaction terms and
produce QUBOs which are not implementable on current hardware.

Thus, modern accurate interatomic potentials require high orders of
interactions while currently constructed hardware is optimized for the
quadratic case. We fill this gap by only considering the total energy returned
by the potential instead of the sum of local interactions. This makes it
possible to use black-box optimization schemes, which iteratively find more
optimal inputs (i.e.\ the structures) to the black-box function (the
potential). In particular, this also opens the possibility to use density
functional theory~\cite{kohn:1965aa} as a source for total energy values.

Various approaches exist to perform materials discovery using black-box
optimization~\cite{terayama:2021aa}. We focus on the ones combining machine
learning principles with quantum
annealers~\cite{kitai:2020aa,koshikawa:2021aa,nusslein:2023aa,schmid:2023aa},
and in particular on Factorization Machines with Quantum Annealing (FMQA) or
Factorization Machines with Annealing (FMA) if performed on classical
devices~\cite{kitai:2020aa,izawa:2021aa,seki:2022aa}. FMQA has been
successfully applied for metamaterials
discovery~\cite{kitai:2020aa,terayama:2021aa} and thus seems particularly apt
for an application to the CSP. FMQA iteratively learns a QUBO representation to
accurately map structures to their black-box calculated energy.

As a QUBO only contains second-order interaction information, we are
effectively learning a pairwise interaction representation for the CSP.\@ It is
generally known that more complicated materials, e.g.\ metals, cannot be
described by pairwise interactions~\cite{thomas:1971aa,daw:1993aa}. Thus, we
cannot expect our FMA approach to produce physically meaningful QUBOs for the
whole spectrum as was the case for the original encoding
in~\cite{couzinie:2024aa}. Instead, as more and more low energy states are
added to the database a second-order approximation of the ground state and its
surrounding states is constructed.

The two main questions we seek to answer thus in this publication are
\begin{enumerate}[(Q1)]
        \item Can we use FMA to sample ground state configurations?
        \item How much information about the non-ground state local minima do the learned QUBOs contain?
\end{enumerate}
To preempt the conclusion our findings are that the algorithm can be used to
efficiently sample ground state configurations for complicated potentials such
as the EAM and that we can quickly infer parts of the relational order of local
minima, while reproducing the actual energy levels requires more intense
calculations.

The rest of the paper is structured as follows. In \cref{sec:methods} we give a
detailed overview of our calculation methods and the materials and potentials
we considered. In \cref{sec:gs_calc} we present FMA calculation results, showing
that we can reproduce the ground state (Q1) for complicated potentials and
\cref{sec:spectrum} discusses the quality of the spectrum of local minima
produced by our approach (Q2). We end by summarising the results in
\cref{sec:conclusions}.

\section{Methods}\label{sec:methods}
In \cref{sub:encoding} we define the functional form of our factorization
machine, in \cref{sub:fma_nagare} the parameter optimization flow we use to
produce ground state configurations and learn good factorization machine
parameters. In \cref{sub:symmetries} we discuss the two main calculation
paradigms we use by accounting or not for symmetry. We end the section by
discussing the three systems and their potentials we use, a Krypton
Lennard-Jones Cluster (\cref{sub:krypton}), which is a pairwise potential and thus exactly learnable by our algorithm. We then consider a Silicon Stillinger-Weber system
(\cref{sub:si}) which is a three-body potential and thus the simplest potential that is not exactly learnable anymore. Finally, we consider a CrFe alloy with an EAM potential (\cref{sub:crfe}) which provides a functionalized approximation for the electron density and thus mimics first principles calculations while keeping the costs low.

\subsection{Encoding}\label{sub:encoding}
Consider a unit cell that is spanned by a given basis $\{\vec{a}_i\}$ with
periodic boundary conditions along a chosen set of basis vectors and a set of
 atom species $\mathcal{S}$. We look at a set of
$N$ lattice points $\emph{X}$ in this unit cell generated by partitioning each
basis vectors into $g$ points and forming the corresponding lattice. The
lattice points have the form $\sum_{i} \frac{k_i}{g} \vec{a}_i$ where $k_i\in \{0,
\ldots, g-1\}$.
Consider a set $b_{x}^{s}$ of binary variables that we define such
that if $b_{x}^{s}=1$ there is an atom of species $s\in \mathcal{S}$ on
$x\in \mathrm{X}$. In~\cite{couzinie:2024aa} a method was introduced that could
encode any interatomic potential with hard cutoff into a HUBO of the form
\begin{align}
        \begin{split}
                &H \\& =\sum_{\substack{x\in \mathrm{X}\\ s\in\mathcal{S}}}
                V^{s}_1(x)b_{x}^{s}
                \\
                   &+\frac{1}{2!}
                   \sideset{}{'}\sum_{\substack{x_1,x_2\in\mathrm{X}\\
                           s_1,s_2\in\mathcal{S}}}
                        V^{s_i,s_j}_2(x_1,
                        x_2)b_{x_1}^{s_1}b_{x_2}^{s_2}+\cdots \\
                   &+\frac{1}{M!}
                   \sideset{}{'}\sum_{\substack{x_1,\ldots, x_M\in X\\
                           s_1,\ldots, s_M\in\mathcal{S}}}
                   V_M^{s_1, \ldots, s_M}(x_1,\ldots,x_M)
                   b_{x_1}^{s_1}\cdots b_{x_M}^{s_M},
        \end{split}
\end{align}
with an appropriately chosen $M$. Using FMA we will approximate this encoding
as Factorization machines~\cite{rendle:2010aa}, which are functions of the
form
\begin{align}\label{eqn:fm}
        f(\mathbf{b}) = \sum_{s\in \mathcal{S}}\sum_{i=1}^{|X|} w_{is} b^s_i
        + \sum_{s_1,s_2\in
        \mathcal{S}}\sum_{i<j}\sum_{n=1}^{k} v_{is_1n}
        v_{js_2n} b^{s_1}_i b^{s_2}_j,
\end{align}
where $k$ is a hyperparameter that governs how many parameters (i.e.\ the set
$\{w_i, v_{isn}\}$) there are in the model and finding binary strings that
minimize $f(\mathbf{b})$ is a QUBO problem. The goal of our FMA approach is finding the parameters such that
the optimal solution to the QUBO problem is the ground state and ideally local
minima correspond to physical local minima.

\subsection{FMA setup}\label{sub:fma_nagare}
\begin{figure}
        \includegraphics[scale=0.65]{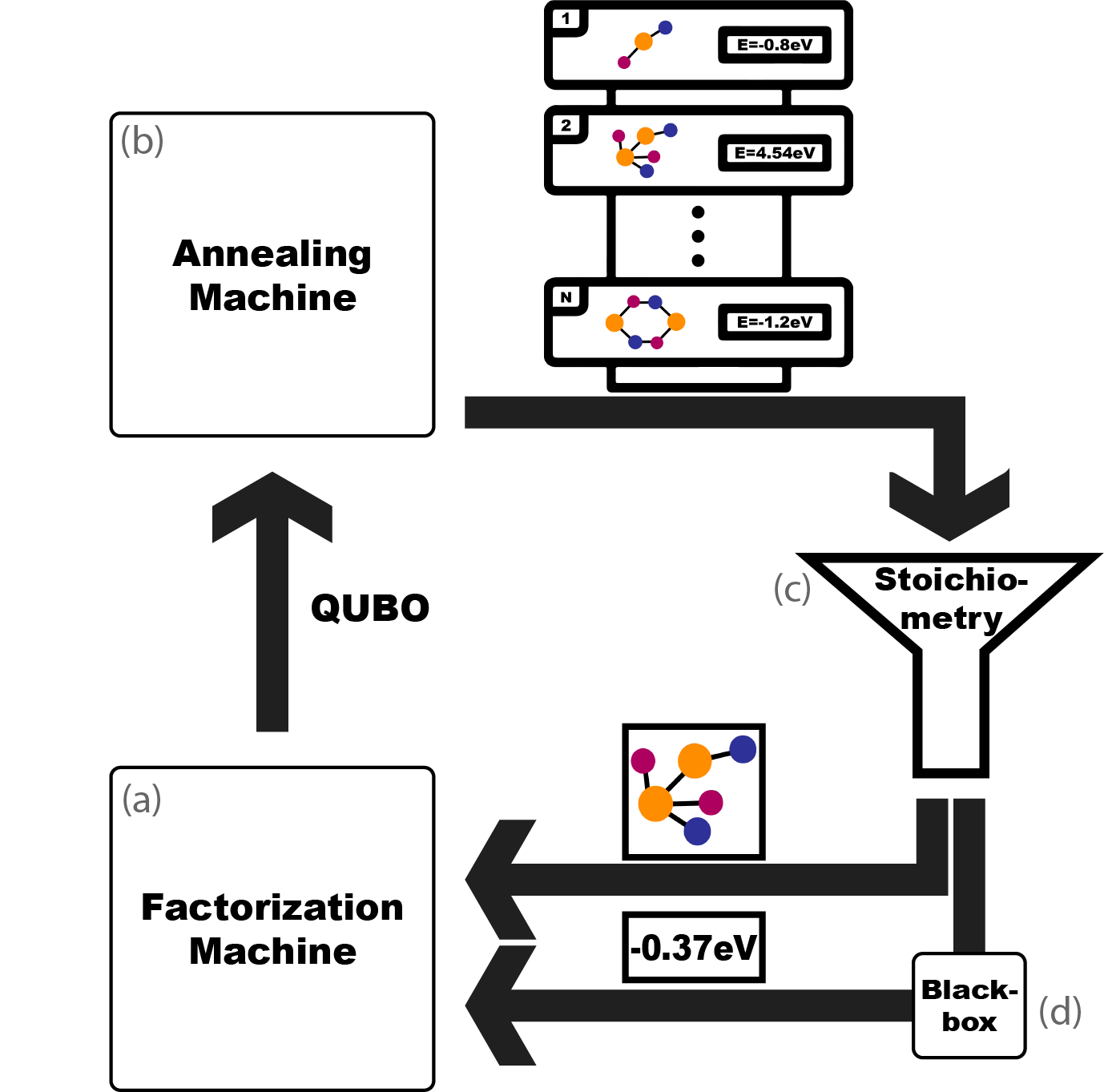}
        \caption{The flowchart of our FMA algorithm. The factorization machine
                (a) is instantiated with an initial dataset and outputs an initial
                QUBO.\@ The annealing machine (b) samples a set number of samples
                from the QUBO.\@ Configurations that do not satisfy the
                stoichiometric constraints are filtered by the stoichiometry
                filter (c). The remaining (lowest energy) configurations are
                evaluated with the black-box function (d) and added to the
                dataset where the cycle repeats starting with the factorization
                machine (a).
                This cycle is repeated for the preset amount of
        iterations.}\label{fig:fma_nagare}
\end{figure}
The general outline of the FMA iteration is given in \cref{fig:fma_nagare}. We
start the FMA calculations by generating an initial dataset comprised of
a preset number of configurations of a system dependent number of atoms on the lattice
together with their respective energies. These configurations are randomly
generated in a way that no two atoms collide and such that the stoichiometric
constraints of the problem are respected. The stoichiometric constraints here refer
to the relative density of the atoms, e.g.\ if we want to optimize for CrFe
configurations we only generate initial configurations that have the same
amount of Cr and Fe atoms. We add a penalty to the energy proportional to how
badly the stoichiometric constraints are broken where the strength of the penalty
$P=20\text{eV}$ is another hyperparameter.

At the sampling stage, we sample $1000$ samples from the resulting QUBO using
simulated annealing~\cite{bertsimas:1993aa} adding a penalty term to the QUBO
that penalises atom collisions, i.e.\ the interaction term between $b_i^{s_1}$
and $b_i^{s_2}$ for two different species $s_1$, $s_2$ on the same lattice site
is increased by $P$. We use the same geometric schedule and bit-flip
neighbourhood as in~\cite{couzinie:2024aa} with maximum temperature $1$,
minimum temperature $0.001$ and $5$ schedule steps. The temperature range is
not problem specific as we normalize the QUBO resulting from the factorization
machine so that individual interaction terms $J_{ij}$ are limited to $[-1, 1]$.
After filtering out the configurations that break the stoichiometric
constraints we transform the binary vectors into configurations readable by the
atomic simulation environment (ASE,~\cite{hjorth-larsen:2017aa}), calculate the
energies using a predefined potential (this is the black-box), add the lowest
energy $5$ configurations to the dataset and start the loop again. The final
output after $50$ iterations is the lowest-energy configuration in the dataset.

\subsection{Accounting for symmetries}\label{sub:symmetries}
In this paper we only consider systems that have periodic boundary conditions
along all three basis directions. Thus, for any particular local minimum
energy there are multiple configurations that are translated or rotated
versions of another. We call calculations with this encoding
\emph{unconstrained}. To account for the translational symmetry we also perform
a \emph{Fixed Zero} calculation in which we fix an atom on the origin, by
setting the corresponding binary variable to $1$ for one of the species and $0$
for the others, reducing the
amount of binary variables in the problem by $|\mathcal{S}|$.

\subsection{Krypton system}\label{sub:krypton}
For the calculation of the potential functions we rely on the Open
Knowledgebase of Interatomic Models (OpenKIM)~\cite{tadmor:2011aa}. In
particular we will look at a three dimensional cubic unit cell of side length
$5.653$\angstrom{} with the Lennard-Jones (LJ) potential parameters due to
Bernades for Krypton~\cite{bernardes:1958aa, elliott:2011aa, tadmor:2011aa,
tadmor:2020aa, tadmor:2020ab}. We discretize the unit cell into an
equipartitioned lattice of $4^3$ nodes by setting $g=4$. The ground state is
given by the FCC configuration and thus encodable on the lattice. The initial
dataset consists of all $64$ single atom Kr$_1$ and all $2016$ two atom Kr$_2$ configurations on the
lattice. For the fixed zero system the initial dataset not only contains the
$63$ Kr$_2$ configurations but also all $1953$ three atom Kr$_3$ configurations. The
energy of the FCC configuration is $-0.119$eV/atom which we take as the zero
energy. We will simply refer to this system as the
Krypton system.

\subsection{Silicon system}\label{sub:si}
We optimize the configuration of pure silicon governed by the Stillinger-Weber
potential parametrisation due to Balamane, Halicioglu and
Tiller~\cite{stillinger:1985aa, balamane:1989aa,balamane:1992aa,
tadmor:2011aa,singh:2021ab,wen:2021aa,tadmor:2020aa,elliott:2011aa}. The unit
cell is cubic with side length $5.44\angstrom$ which we again discretize into
$4^3$ nodes by setting $g=4$. The well-known Si$_8$ crystal ground state
configuration is then encodable on the lattice. The energy of the ground state
is $-4.630$eV/atom which we set as the $0$ energy for this system. The initial
dataset consists of $8000$ randomly sampled Si$_8$ configurations on the
lattice.

\subsection{CrFe system}\label{sub:crfe}
We optimize the configuration of a chrome-iron alloy governed by the modified
embedded atom-method potential parametrisation due to Lee, Shim and
Park~\cite{afshar:2023aa,lee:2001aa,lee:2023aa,tadmor:2020aa,elliott:2011aa}.
We use an orthorhombic unit cell with side length $2.92310\angstrom$,
$3.88108\angstrom$, $4.02297\angstrom$ found on Materials
Project~\cite{jain:2013aa}\footnote{Data retrieved from the Materials Project
for CrFe (mp-1226211) from database version v2023.11.1.}. We discretize this
unit cell again into $4^3$ nodes by setting $g=4$, so that the problem has
$2\cdot 64=128$ bits as we have two atom species. The reference
structure is a Cr$_2$Fe$_2$
configuration that has $-4.088$eV/atom as the ground state energy which we set
as the $0$ energy. Our initial datasets consist of
$8000$ randomly sampled Cr$_2$Fe$_2$ configurations on the lattice.

\section{Finding the ground state}\label{sec:gs_calc}
\subsection{Setup}
We perform calculations for two sets of parameters, one that we call the
\emph{quick settings} and one we call \emph{accurate settings}. The idea is
that to find optimal crystal configurations speed is more important than having
accurate QUBOs. To investigate how well we can approximate our HUBO as a QUBO
we will use the more time intensive accurate settings.

In both settings we initialize the factorization machines with random weight
and use the AdamW method to optimize them. In the quick settings we set $k=20$
use a learning rate of $0.08$ and learn for $10$ epochs for a total of $50$
iterations. This calculation aims at quickly sampling the iteratively learned
QUBO to find the ground state.

In the accurate settings we set $k=70$ for the Krypton and Silicon systems and
$k=140$ for the CrFe system. We set the learning rate to $0.01$ and learn for
$2500$ epochs and a total of $20$ iterations. We determined these settings by
increasing the epoch number, and $k$, and decreasing the learning rate until
the Krypton system (whose potential is pairwise, and thus perfectly
representable as a QUBO) was perfectly learned by the FMA algorithm in one
iteration.

To account for the randomness in the initial dataset and FMA initialization we
perform $30$ independent FMA runs for each system and extract the QUBO
producing the lowest energy sample from each run. Note that the \emph{learned energy} (i.e.\
the energy as given by the QUBO) that gets
sampled does not match the \emph{reference energy} (i.e.\ the energy as calculated by
the black-box potential function), and we determine the lowest energy
configuration by considering its reference energy and not its learned energy.
We use the word \emph{learned} to refer to the results as produced
by the FMA algorithm and \emph{reference} to refer to the true values as given
by the potential function. To judge how good
the algorithm is at finding the reference ground state we look at how many runs
return the reference ground state and how many iterations it takes on average
to recover the reference ground state. We do this separately for the
unconstrained (\cref{tab:improv_rate}) and fixed zero (\cref{tab:gs_iters})
systems.

\subsection{Results \& Discussion}
\begin{figure}
        \includegraphics[scale=0.6]{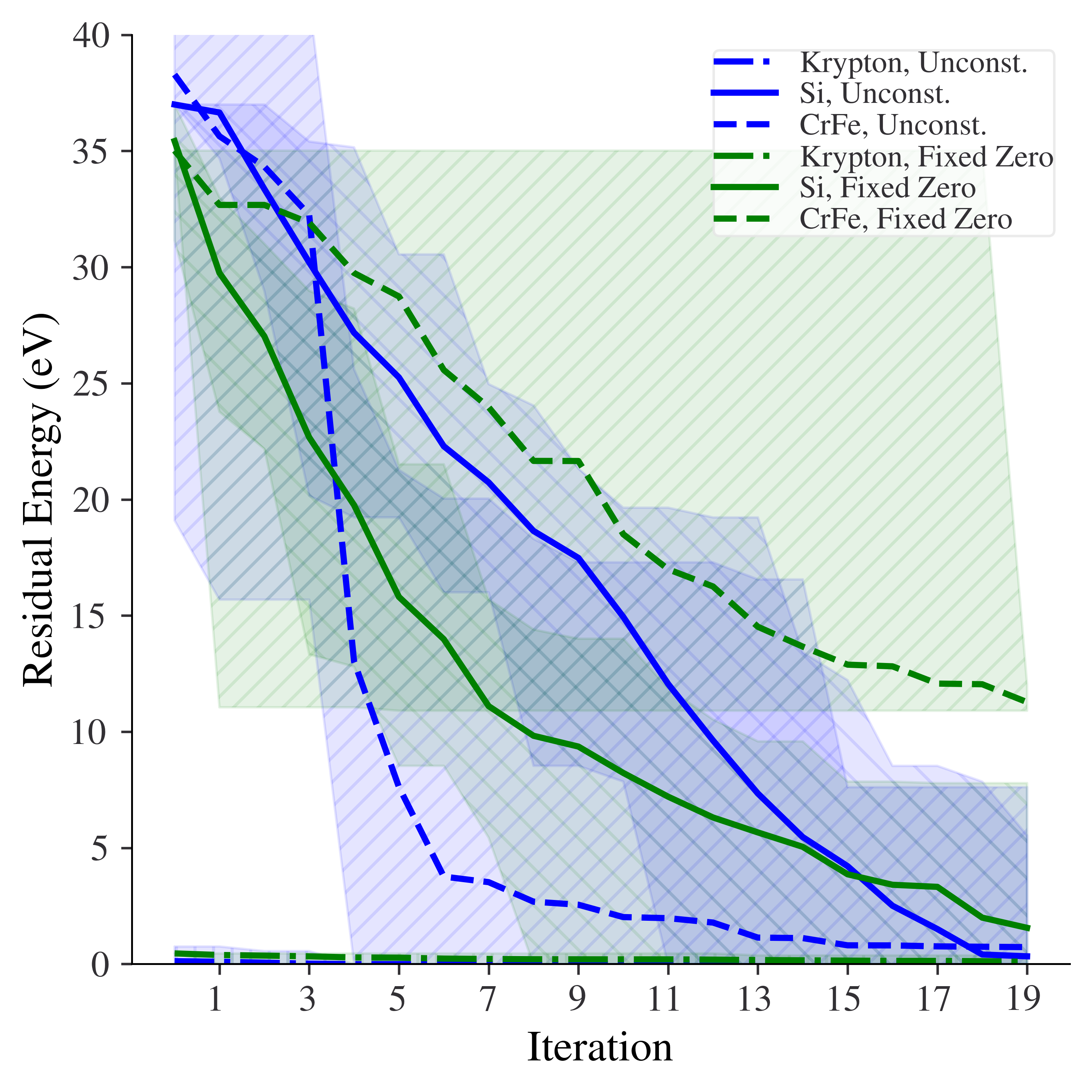}
        \caption{Cumulative minimum over 30 FMA runs for the quick setting plotted agains the iteration until which the cumulative minimum is taken. The solid, dashed and dash-dotted lines are the mean cumulative minimum and the shaded areas the corresponding minimum and maximum cumulative minimum until that iteration. Silicon is in solid, CrFe in dashed, Krypton in dash-dotted. Unconstrained calculations are in blue and fixed zero calculations in green. Bottom-left to top-right shaded areas are for CrFe and the inverse orientation for Silicon. 
         Horizontally dashed lines are for Krypton.\label{fig:cum_min}}
\end{figure}

\begin{table}
        \begin{tabular}{lcccc}
                 & \multicolumn{2}{c}{Quick} & \multicolumn{2}{c}{Accurate} \\
                System & Count & Iterations & Count & Iterations \\
                \hline
                Krypton & $30$ & $1.7 \pm 0.23$ & $30$ & $1\pm 0$ \\
                Si & $30$ & $17.0 \pm 0.36$ & $30$ & $7.0 \pm 0.57$ \\
                CrFe & $24$ & $17.9 \pm 2.11$ & $28$ & $11.9 \pm 0.58$ \\
        \end{tabular}
\caption{\label{tab:improv_rate} The number of runs (of a total of $30$) where
the ground state is found during one of the FMA iterations for the
unconstrained system and the average amount of iterations with standard
deviation for the first iteration at which the ground state was sampled.
The Krypton values for the accurate settings are $1\pm 0$ as the parameters are chosen such
that the exact representation is learned  in one iteration.}
\end{table}
We see in \cref{tab:improv_rate} that for the unconstrained system we are able
to sample the reference ground state in the quick settings in all iterations for the Krypton
and Si systems and in $24$ of $30$ iterations for the CrFe system, which is
improved to $28$ in the accurate settings. We also see that, as expected, the
accurate settings
requires less iterations until the reference ground state is sampled, which
likely stems from the higher number of performed epochs leading to a more accurate
second order approximation of the dataset and thus to more high-quality
annealing sample. For the fixed zero calculations results (\cref{tab:gs_iters})
we see that for the quick settings in the Krypton system we only find the
ground state for $25$ iterations taking $10.7$ iterations on average as opposed
to $1.7$ for the unconstrained system. For the silicon system the iteration
number is slightly increased and in the accurate settings we only manage to
find the ground state in $25$ out of $30$ runs. For the CrFe system with fixed
zeros we do not manage to sample the ground state in any iteration. This might
be due to a local minimum consistently having a lower learned energy than the
reference ground state (see for example
\cref{fig:best_kendall} in the next section).
\begin{table}[]
        \begin{tabular}{lcccc}
                 & \multicolumn{2}{c}{Quick} & \multicolumn{2}{c}{Accurate} \\
                System & Count & Iterations & Count & Iterations \\
                \hline
                Krypton & $25$ & $10.7 \pm 1.42$ & $30$ & $1\pm 0$ \\
                Si & $30$ & $18.4 \pm 1.14$ & $25$ & $8.3 \pm 0.95$ \\
                CrFe & $0$ & N/A & 0 & N/A \\
        \end{tabular}
        \caption{\label{tab:gs_iters}Same as \cref{tab:improv_rate} for the
        fixed zero system.}
\end{table}
In \cref{fig:cum_min} we see the cumulative minimum over the 30 FMA runs for the quick setting. It is easily observable that the Fixed zero calculations (green) perform worse than the unconstrained ones (blue) for the non-exactly learnable systems of Si and FeCr. Further, while the average iteration to find the ground state was similar in \cref{tab:gs_iters} we see that unconstrained CrFe tends to have a lower cumulative minimum and quickly goes to a residual energy of around 3-4 eV and then slowly decreases while Silicon has a more linear deterioriation. For the fixed zero calcuations we see that CrFe slowly decreases to a non-zero residual energy, which is reflected in the fact that we do not sample the ground state while the Silicon calculation is roughly comparable to the unconstrained case. In total we see that fixing the zero gives comparable or worse performance.

It is expected that the Krypton system outperforms both the Silicon and the
CrFe system at the quick settings, as the reference LJ potential is a pairwise potential and hence
representable as a QUBO as done in~\cite{gusev:2023aa,couzinie:2024aa} and
recovered for the accurate settings. Thus,
if the parameter for the interaction between atoms on two sites $x$ and $y$ of
the lattice is correctly estimated to reproduce the energy of a configuration
in the dataset, it will be correctly estimated for any configuration in the
dataset and should thus be correctly reflected in the learned potential. For
the higher order potentials the energy contribution of a pair of atoms is much
harder to estimate as the contribution in the reference potential might vary
drastically depending on the surrounding configuration, e.g.\ the interaction
between $x$ and $y$ depends on the angles it forms with other atoms
neighbouring them.

To summarize the results,
considering the unconstrained system we see that our FMQA setup is able to
efficiently sample the ground state configurations of arbitrary potentials,
even if multiple iterations are recommended to ensure that the ground state is
found (see CrFe results in \cref{tab:improv_rate}). As opposed to previous
findings~\cite{seki:2022aa}, we do not see evidence that reducing the bit
number necessarily has a favorable impact on the difficulty of the problem.

\section{Reproduction of order of local minima}\label{sec:spectrum}

\subsection{Setup}
We consider the quick and accurate settings for the local minima reproduction.

We extract the set of all local minima of the reference potential represented
on our lattice. We define a state on the lattice as a reference local minimum
if it has an atom density of that of the reference ground state or lower,
satisfies the stoichiometric constraints, has an absolute force on the atoms of
$0$ and if its energy is lower than $0$. Note that in our encoding we provide
no information on the force and thus, due to the discretization, there might be
other meta-stable local minima when constraining the system to the lattice,
nonetheless we only consider the force-free local minima as given by the
reference potential, as these are the physical local minima.

We judge the reproduction of the reference spectrum using the relational order
of the learned energy. This is because as long as the learned energy levels
have the correct relational order, their reference value is not as important,
as it is usually inexpensive to recalculate the energies of the local minima if
they are correctly identified.

We do this using
the Kendall rank coefficient~\cite{kendall:1938aa} a real number between $-1$
and $1$. In our case, a value of $1$ indicates that the learned local minima
energy levels have the same order as the reference ones, and a value of $-1$
that the order is reversed while $0$ indicates no order. Ideally, FMA would
produce QUBOs that have a value of $1$.

For each set of calculation parameters we consider the iteration that produces
the highest Kendall rank coefficient QUBO.

\subsection{Results \& Discussion}

\begin{table}
        \begin{tabular}{lcccc}
                 & \multicolumn{2}{c}{Quick} & \multicolumn{2}{c}{Accurate} \\
                 & Unconstrained & Fixed Zero & Unconstrained & Fixed zero\\
                \hline
                Krypton & $0.32$ & $0.72$ & $1$ & $1$ \\
                Si & $0.32$ & $0.44$ & $0.26$ & $0.45$ \\
                CrFe & $0.48$ & $0.69$ & $0.41$ & $0.68$ \\
        \end{tabular}
        \caption{\label{tab:kendall} The results for the Kendall coefficient.}
\end{table}
The results for the Kendall coefficients for the various systems are given in
\cref{tab:kendall}. We see that generally the fixed zero system improves the
Kendall coefficient and that the CrFe system outperforms the Silicon system. To
understand this, consider that for the Silicon system we have $7056$ ($482$)
local minima and for the CrFe system we have $1856$ ($52$) with the fixed zero
system value in brackets. Since FMQA performs a quadratic approximation of the
potential energy landscape around the states in the database, having fewer
local minima that get sampled makes it easier to represent these accurately.
This is particularly apparent for the $0.72$ value of fixed zero Krypton in
which only $20$ local minima are considered.

\begin{figure}
        \includegraphics[scale=0.7]{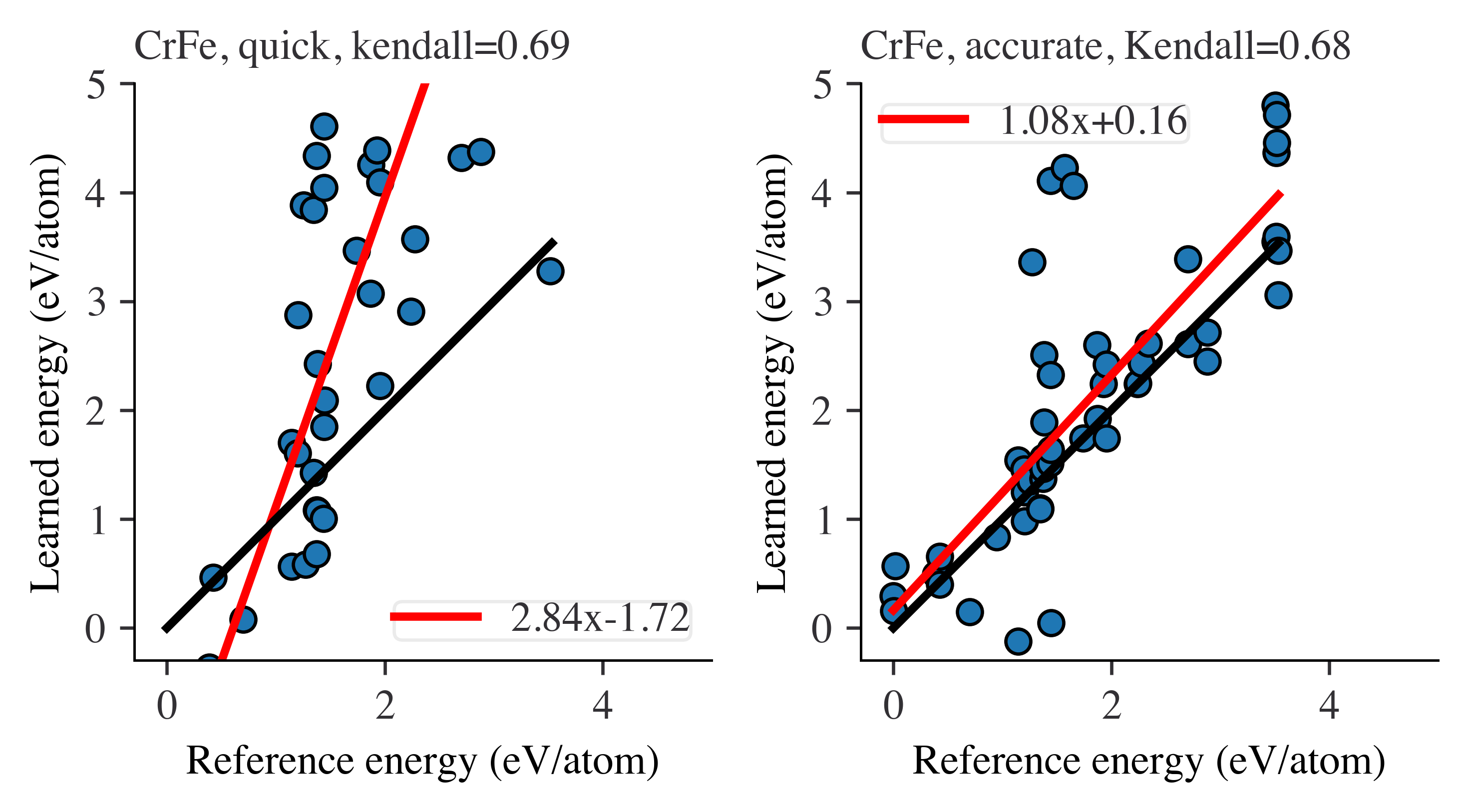}
        \caption{Learned energy levels as blue dots for the fixed zero CrFe
                system for the quick settings (left) and accurate settings
                (right) plotted against the energy for the reference potential.
                The black diagonal corresponds to the reference values and the
                red line to a linear regression resulting in a line
                $2.84x-1.72$ for the quick settings and $1.08x+0.16$ for the accurate
                settings. The Kendall rank coefficient is $0.69$ and $0.68$
                respectively.\label{fig:best_kendall}}
\end{figure}

The quick settings also tend to slightly outperform the accurate settings. A
possible explanation is given by \cref{fig:best_kendall} in which the learned
energy levels (blue dots) are plotted against their reference values for the
quick (left) and accurate (right) settings of the CrFe fixed zero system. It is
apparent that in the accurate settings, we more accurately reflect the actual
energy levels of the potential. The fit is not close to chemical accuracy, which is easily explained by considering that we are approximating a many-body potential
with a quadratic function. In particular this shows how well FMQA can
quadratize a HUBO problem into a QUBO. The quick settings on the other hand do not
represent the actual values of the energy levels well but their relational
order as there is more distance between the various local minima.

Notice also, that in the shown accurate settings the ground state of the learned energies and reference energies are not equal and we can sample the ground state only because we take 1000 samples from the QUBO at every iteration. This reliance on a high sample number could be problematic if our black-box is a first principle based calculator with high computational costs.

These results hint at the possibility that to learn relational orders an
exploratory learning phase such as in the quick settings is enough, and even
beneficial as local minima whose reference value is close are learned with
values far apart. The accurate settings lend themselves more to learn a
quadratic representation of the potential.

\section{Conclusions}\label{sec:conclusions}
In conclusion, this publication explores the application of Factorization
Machines with Quantum Annealing (FMQA) to the crystal structure problem (CSP)
in materials science. The CSP involves determining the optimal arrangement of
atoms in a material based on its chemical composition. The study investigates
the capability of FMQA to efficiently sample ground state configurations and
reproduce the spectrum of local minima energy levels.

The results show that FMQA can be used to quickly sample ground state
configurations of various potential forms including three-body Stillinger-Weber
potentials and EAM potentials. 

We show that using calculation settings that quickly finish the relational order between local minima is
well represented in particular when accounting for translational symmetry by
fixing an atom on the origin. Factorization machines are known to lend themselves well for machine learned ranking~\cite{rendle:2010aa} this work marks a first exploration into how well the FMA algorithm can be applied on the same problem. Using more time-intensive accurate settings we further
show that the local minima of a metal alloy using an EAM
potential is well approximated by the quadratic QUBO matrix. While the accuracy
is not enough to serve as an interatomic potential, we have shown the potential
of FMQA as a quadratization scheme for HUBO problems.

As opposed to previous findings~\cite{seki:2022aa}, we did not observe that reducing the search space improves the quality of the results. The results in \cite{seki:2022aa} are not directly applicable here, but our findings still raise the need for further investigation into the relation of search space size and FMQA performance.

There are several ways the FMQA algorithm could be improved to potentially
provide more faithful approximated potentials. An obvious improvement is going
to higher-order factorization machines~\cite{rendle_higher_order,blondel:2016aa}, though this
suffers from the same problem as the higher-order encoding presented
in~\cite{couzinie:2024aa} in that sampling HUBOs is not efficient on current
hardware. Recently high-quality non-parametric potentials 3-body potentials
have been proposed~\cite{glielmo:2017aa,vandermause:2020aa} and so it might not be necessary
to go to higher-order but rather add non-linearities for example by replacing
the dot product in factorization machines with kernels~\cite{bishop:2006aa}.

The study highlights the importance of considering the nature of interatomic
potentials and system complexity when applying quantum-enhanced algorithms to
materials science problems. The findings contribute to the ongoing exploration
of quantum computing methodologies for addressing challenges in materials
discovery and crystal structure determination.

\begin{acknowledgments}
This work was supported by JSPS KAKENHI as ``Grant-in-Aid for Scientific
Research(A)'' Grant Number 21H04553 and ``Grant-in-Aid for Scientific Research(S)'' JP23H05447. 
This paper is partially based on results obtained from a project, JPNP23003, commissioned by the New Energy and Industrial Technology Development Organization (NEDO).
This work was supported by the Center of Innovations for Sustainable Quantum AI (JST Grant Number JPMJPF2221). 
This work was partially supported by the Council for Science, Technology, and Innovation (CSTI) through the Cross-ministerial Strategic Innovation Promotion Program (SIP), ``Promoting the application of advanced quantum technology platforms to social issues'' (Funding agency: QST).
S.~T. wishes to express their gratitude to the World Premier International Research Center Initiative (WPI), MEXT, Japan, for their support of the Human Biology-Microbiome-Quantum Research Center (Bio2Q).

The computation in this work has been done
using the supercomputer provided by Supercomputer Center at the Institute for
Solid State Physics at the University of Tokyo.
\end{acknowledgments}
\bibliography{quantum_computing}
\end{document}